\documentclass[twocolumn,english,aps,pra]{revtex4}
\usepackage[T1]{fontenc}
\usepackage[latin9]{inputenc}
\usepackage{babel}
\usepackage{amssymb}
\usepackage{graphicx}
\usepackage[unicode=true]
 {hyperref}
\usepackage{breakurl}

\makeatletter
\@ifundefined{textcolor}{}
{%
 \definecolor{BLACK}{gray}{0}
 \definecolor{WHITE}{gray}{1}
 \definecolor{RED}{rgb}{1,0,0}
 \definecolor{GREEN}{rgb}{0,1,0}
 \definecolor{BLUE}{rgb}{0,0,1}
 \definecolor{CYAN}{cmyk}{1,0,0,0}
 \definecolor{MAGENTA}{cmyk}{0,1,0,0}
 \definecolor{YELLOW}{cmyk}{0,0,1,0}
}

\makeatother

\begin{document}

\title{Genuine multipartite system-environment correlations in decoherent
dynamics}

\author{Jonas Maziero}

\email{jonasmaziero@gmail.com}

\selectlanguage{english}%

\author{F\'{a}bio M. Zimmer}

\email{fabiozimmer@gmail.com}

\selectlanguage{english}%

\address{Departamento de F\'{i}sica, Universidade Federal de Santa Maria, 97105-900,
Santa Maria, RS, Brazil}
\begin{abstract}
We propose relative entropy-based quantifiers for genuine multipartite total, quantum, and classical correlations. These correlation measures are applied to investigate the generation of genuine multiparticle correlations in decoherent dynamics induced by the interaction of two qubits with local-independent environments. We consider amplitude- and phase-damping channels and compare their capabilities to spread information through the creation of many-body correlations. We identify changes in behavior for the genuine $4$- and $3$-partite total correlations
and show that, contrary to amplitude environments, phase-noise channels
transform the bipartite correlation initially shared between the qubits
into genuine multiparticle system-environment correlations.
\end{abstract}
\maketitle

\section{Introduction}

The nonlocality \cite{Aspect-Review,Mermin-RMP}, nonseparability
\cite{Horodecki-RMP,Guhne-PR}, and quantumness \cite{Celeri-Review,Modi-Review}
of correlations in composite systems are currently recognized as important
ingredients for the efficiency of protocols in quantum information
science. Substantial progress has been achieved latterly regarding
the characterization, identification, and quantification of total,
classical, and quantum correlations \cite{Groisman-PRA,Henderson-JPA,Ollivier-QD,Horodecki-Defict,Piani-NLB,Cavalcanti-SMerg,Madhok-SMerg,Ferraro,Zurek-MD,Adesso-CV,Giorda-CV,Cornelio-EIrrev,Streltsov-DEM,Maziero-Wit,Aguilar-WPRL}.
The nonclassicality originated from local indistinguishability of
quantum states is the more recent paradigm for analyzing correlations.
This kind of correlation was studied mostly in the case of bipartite
systems, with novel and interesting results already obtained. By its
turn, the investigation of (genuine) multipartite quantum and classical
correlations has been receiving its deserved attention only very recently
\cite{kaszlikowski,walczak,grudka-arxiv,Saitoh-IJQI,Rulli-Sarandy-GQD,Chakrabarty-MQD,Okrasa-EPL,Parashar-Rana,Campbell-IJQI,Giorgi-GMTC-XY,Byczuk-PRL,Ranzan-Dynamics-GQD,Comment-Paper-Giorgi,Square-D-&-Monogamy,Average-GMQD,Mahdian-MQD-Dec,Xu-GQD,GMQD-Detection,Saguia-Wit,Modi-reldis,Giorgi-genuine,Bennett-genuine}.
In this article we are interested in this last scenario. We shall
define measures for genuine $n$-partite correlations (Section \ref{sec:quantifiers})
and study how these correlations develop during decoherent system-environment
dynamics (Section \ref{sec:dynamics}).

\subsection{The decoherence process}
The thorough investigation of the decoherence process \cite{Zurek-RMP,Breuer-Petruccione,Schlosshauer}
is an important issue to be addressed towards large scale implementations
of, for instance, quantum computers, quantum simulators, and communication
protocols. The decoherence phenomenon is a result of the inevitable
interaction between a quantum system and its surroundings. Let us
consider a system $s$ and its environment $E$ in an initial product
state $\rho_{sE}=\rho_{s}\otimes|0_{E}\rangle\langle0_{E}|$, where $|0_{E}\rangle$ is the vacuum state of the environment. The dynamics of the whole closed system is unitary, i.e., $\rho_{sE}(t)=U_{sE}(t)\rho_{sE}U_{sE}^{\dagger}(t)$
with $U_{sE}(t)U_{sE}^{\dagger}(t)=\mathbb{I}$, where $\mathbb{I}$
is the identity operator. Then, by tracing out the environmental variables
we write down the system's evolved state in the Kraus or operator-sum
representation: $\mathcal{E}(\rho_{s})=\sum_{i}K_{i}\rho_{s}K_{i}^{\dagger}$,
where $K_{i}:=\langle i_{E}|U_{sE}(t)|0_{E}\rangle$ are linear operators
on the state space of the system, $\mathcal{H}_{s}$, such that $\sum_{i}K_{i}^{\dagger}K_{i}\le\mathbb{I}$ and $\{|i_{E}\rangle\}$ is a basis for the environment with $i$ specifying number of excitations that is distributed in all its modes.
We observe that the quantum operation $\mathcal{E}$ can be used to describe general
transformations between quantum states \cite{Nielsen-Chuang,Preskill}.

One can verify that the dynamical map for the evolution of the system-environment
state:
\begin{equation}
U_{sE}|\psi_{s}\rangle\otimes|0_{E}\rangle=\sum_{i}K_{i}|\psi_{s}\rangle\otimes|i_{E}\rangle,
\end{equation}
leads to the same motion equation for the system state as shown above.
Thus, we can obtain the Kraus' operators describing the noise channel
acting on the system using a phenomenological approach \cite{Almeida-POA,Diogo-Dec-Control}
or via quantum process tomography \cite{Chuang.Nielsen-PT,Vandersypen.Chuang-RMP}
and use this information to investigate, for example, the system-environment
correlations, without worry about the usually complicated structure
of the environment. 

\subsection{A partial classification of quantum states}
\label{state-classification}

A possible, partial, classification of multipartite quantum states
concerning its correlations, or with regard to the operations needed
to generate such correlations, can be introduced as follows. For an
$n$-partite system prepared in a product state vector $|\psi_{\mathrm{init}}\rangle=|\psi_{01}\cdots\psi_{0n}\rangle,$
any uncorrelated $n$-partite product state of this system,
\begin{equation}
\rho_{1\cdots n}^{p}=\rho_{1}\otimes\cdots\otimes\rho_{n},
\end{equation}
can be created by means of local quantum operations (LQO). The sub-index
$is$ in $|\psi_{\mathrm{init}}\rangle$ specifies the state and subsystem,
respectively, and we use throughout this article the notation $|\psi_{01}\cdots\psi_{0n}\rangle:=|\psi_{01}\rangle\otimes\cdots\otimes|\psi_{0n}\rangle$.

Starting with the system in state $|\psi_{\mathrm{init}}\rangle$,
any $n$-partite classically-correlated state can be prepared via
local classical operations (LCO) coordinated by the exchange of classical
communication (CC) and has the form:
\begin{equation}
\rho_{1\cdots n}^{c}=\sum_{i1,\cdots,in}p_{i1,\cdots,in}|\psi_{i1}\cdots\psi_{in}\rangle\langle\psi_{i1}\cdots\psi_{in}|,
\end{equation}
where the states $\{|\psi_{is}\rangle\}\in\mathcal{H}_{s}$ form a
complete ($\sum_{is}|\psi_{is}\rangle\langle\psi_{is}|=\mathbb{I}_{s}$)
orthonormal ($\langle\psi_{is}|\psi_{js}\rangle=\delta_{ij}$), and
therefore distinguishable, basis for the subsystem $s$. Above, by
LCO we mean (complete) transformations between the pointer basis states
\cite{Zurek-RMP}. It can be seen that if the probability distribution
in the state $\rho_{1\cdots n}^{c}$ factorizes, i.e., $p_{i1,\cdots,in}=p_{i1}\cdots p_{in}$,
there is no correlation at all in the system, that is to say, it is an $n$-partite
product state. 

An $n$-partite separable but quantumly correlated state
needs LQO and CC to be generated from $|\psi_{\mathrm{init}}\rangle$.
This kind of state has the general form
\begin{equation}
\rho_{1\cdots n}^{s}={\textstyle \sum_{i}}p_{i}\rho_{i1}\otimes\cdots\otimes\rho_{in},
\end{equation}
with $\{p_{i}\}$ being a joint probability distribution and $\{\rho_{is}\}$
being noncommuting density operators. We note that if the density
operators $\rho_{is}$ commute for different $i$, then the state
$\rho_{1\cdots n}^{s}$ is an $n$-partite classically-correlated state.
If in addition the probability distribution $\{p_{i}\}$ factorizes,
then $\rho_{1\cdots n}^{s}$ is an $n$-partite product state. 

The $n$-partite entangled, or non-separable, states $\rho_{1\cdots n}^{e}$
cannot be prepared locally, requiring direct or mediated interaction
for its generation. One famous example of entangled state is the GHZ
state \cite{GHZ,GHZ-E}:
\begin{equation}
2^{-1/2}(|0_{1}\cdots0_{n}\rangle+|1_{1}\cdots1_{n}\rangle),
\end{equation}
where $\{|0_{s}\rangle,|1_{s}\rangle\}$ is the
one-qubit computational basis.

\subsection{Relative entropy-based measures of correlation}
\label{subsec-rebq}

Considering the operational state classification of the last subsection and with the aim of quantifying the different kinds of correlation in an unified manner, Modi \emph{et al.} \cite{Modi-reldis} introduced measures of correlation using the relative entropy \cite{Vedral-RE},
\begin{equation}
S(\rho||\sigma)=\mathrm{tr}(\rho(\log_{2}\rho-\log_{2}\sigma)),
\end{equation}
to estimate the ``distance'' between two states $\rho$ and $\sigma$.
The total correlation in an $n$-partite state $\rho_{1\cdots n}$
is quantified by how distinguishable or how distant it is from an
uncorrelated $n$-partite product state \cite{Modi-reldis}, i.e.,
\begin{eqnarray}
I(\rho_{1\cdots n}) & = & \min_{\rho_{1\cdots n}^{p}}S(\rho_{1\cdots n}||\rho_{1\cdots n}^{p})\nonumber \\
 & = & S(\rho_{1\cdots n}||\rho_{1}\otimes\cdots\otimes\rho_{n}).
\end{eqnarray}
The last equality was established in Ref. \cite{Modi-reldis} and
shows that the closest $n$-partite product state of any state $\rho_{1\cdots n}$
is obtained from its marginal states in the product form. Recalling
the state classification presented in the previous paragraph, the
quantum part of the correlation in $\rho_{1\cdots n}$ can be defined
as its minimal distance from $n$-partite classically-correlated states \cite{Modi-reldis}:
\begin{eqnarray}
Q(\rho_{1\cdots n}) & = & \min_{\rho_{1\cdots n}^{c}}S(\rho_{1\cdots n}||\rho_{1\cdots n}^{c})\nonumber \\
 & = & S(\rho_{1\cdots n}||\chi_{1\cdots n}^{\rho}),
\end{eqnarray}
with $\chi_{1\cdots n}^{\rho}=\sum_{i1,\cdots,in}p_{i1,\cdots,in}|\psi_{i1}\cdots\psi_{in}\rangle\langle\psi_{i1}\cdots\psi_{in}|$
and $p_{i1,\cdots,in}=\langle\psi_{i1}\cdots\psi_{in}|\rho_{1\cdots n}|\psi_{i1}\cdots\psi_{in}\rangle$
\cite{Modi-reldis}. In the second equality for $Q$, we leave implicit
the minimization over local basis needed to find the closest classically-correlated
state $\chi_{1\cdots n}^{\rho}$. Finally, the classical part of the
correlations in $\rho_{1\cdots n}$ is given by the total correlation
of $\chi_{1\cdots n}^{\rho}$ \cite{Modi-reldis}:
\begin{eqnarray}
C(\rho_{1\cdots n}) & = & \min_{\rho_{1\cdots n}^{p}}S(\chi_{1\cdots n}^{\rho}||\rho_{1\cdots n}^{p}),
\end{eqnarray}
and quantifies how distant $\chi_{1\cdots n}^{\rho}$ is from being an uncorrelated state.

\subsection{Genuine multipartite correlations}
\label{subsec-gmc}

Though the state classification and the correlation quantifiers presented
above are very useful in several contexts, they are restricted in
a certain sense because they classify the system state without considering
the possibility of a more complex distribution for the correlations.
For instance, in studying multipartite spin systems, which may have
a far more intricate density operator with groups or cluster of spins
in different families of states, it would be desirable to generalize
the results presented in the last two paragraphs in order to include
and quantify such complexity. As we will discuss in the sequence,
the concept of genuine multipartite correlation fits well for the
study of these more general scenarios and can be utilized as the starting
point to define generalizations of the correlation quantifiers
discussed above. 

Addressing this subject, Bennett \emph{et al.} (see
Ref. \cite{Bennett-genuine})\emph{ }postulated that if an $n$-partite
state has no genuine $n$-partite correlation, then we cannot create
genuine $n$-partite correlation by adding a subsystem in a product
state or by performing trace non-increasing local quantum operations.
Also, we cannot create genuine ($n+1$)-partite correlation by splitting
a subsystem in two. They proved that the following definitions satisfy
these requirements. An $n$-partite state has genuine $n$-partite
correlation only if it is non-product under any bipartite cut of the
system. For $k\le n$, an $n$-partite state has genuine $k$-partite
correlation only if there exists a subset of $k$ subsystems presenting
genuine $k$-partite correlation. Bennett and collaborators also defined
the degree of correlation of an $n$-partite state as the maximum number
$k$ of subsystems possessing genuine $k$-partite correlation.

\section{Quantifiers for Genuine Multipartite Correlations}

\label{sec:quantifiers}

The investigation of the different types of correlation presented in quantum states is one of the main problems in quantum information science. Recently the identification and quantification of genuine multipartite correlations (which are those correlations we cannot account looking only to a part of the system) has been receiving its first studies. In this section we shall introduce definitions and quantifiers for genuine multipartite correlations by using the concepts reviewed in Section \ref{subsec-gmc} to generalize the measures discussed in Section \ref{subsec-rebq}.
 
From the definition given in Section \ref{subsec-gmc}, it follows that an $n$-partite state
has genuine $n$-partite correlation only if it is non-product under
any bipartite cut of the system, i.e., if $\rho_{1\cdots n}\ne\rho_{c_{1}}\otimes\rho_{c_{2}},$
where $c_{1}\mbox{ }(c_{2})$ indicates a group with a number $n_{1}\mbox{ }(n_{2})$
of subsystems, and $n_{1}+n_{2}=n$. In other words, if $\rho_{1\cdots n}$
possesses genuine $n$-partite correlation then there does not exist
two completely uncorrelated clusters of particles in the system. Thus,
a measure for \textit{genuine $n$-partite total correlation} of an $n$-partite
state can be logically defined as
\begin{equation}
I_{n}(\rho_{1\cdots n}):=\min_{(c_{1},c_{2})}S(\rho_{1\cdots n}||\rho_{c_{1}}\otimes\rho_{c_{2}}),
\end{equation}
where the minimum is taken over all possible bi-partitions $(c_{1},c_{2})$
of the system. Therefore the quantity $I_{n}$ measures the minimum
distance between $\rho_{1\cdots n}$ and states that do not possess
genuine $n$-partite correlation. 

Starting from a possible generalization of Shannon's mutual information for tripartite systems, the quantifier for genuine $n$-partite total correlation $I_{n}(\rho_{1\cdots n})$ was also noticed by Giorgi
and colleagues in Ref. \cite{Giorgi-genuine}. They also defined measures for genuine classical and quantum correlations using a quantum discord-like approach, i.e., assuming additivity of mutual information in classical
and quantum correlations. As shown in Ref. \cite{Comment-Paper-Giorgi}, the complicated alternative definitions presented in Ref. \cite{Giorgi-genuine} for genuine $n$-partite quantum and classical correlation of $n$-partite states do not coincide in general.
In sequence of this article, we shall start from a direct extension of the operational classification of states discussed in Section \ref{state-classification} and use relative entropy-based distinguishability measures to define simple and general correlation quantifiers for genuine multipartite correlations that are valid for any number of subsystems and for any dimension of its Hilbert spaces. It is also worthwhile to mention that, although we choose the relative entropy as a measure of distance, our quantifiers for correlation can be defined in a similar manner using other distance measures.

Regarding total correlations, we can also look to the structure of the system's correlation by defining a measure for \textit{genuine $k$-partite total correlation} of an $n$-partite state:
\begin{equation}
I_{k}(\rho_{1\cdots n}):=\max_{\rho_{k}}I_{k}(\rho_{k}).
\end{equation}
In the last equation, the maximum is taken over all states $\rho_{k}$ comprising $k$ subsystems and aims to identify the most correlated $k$-partite  group of particles in the system. For example, if $k=n-1$, $I_{n-1}(\rho_{1\cdots n})$ is the maximal genuine $(n-1)$-partite total correlation of the states obtained by tracing out one subsystem of $\rho_{1\cdots n}$. 

This last quantifier, $I_{k}(\rho_{1\cdots n})$, can be applied to study the \textit{degree of correlation} (as defined in Section \ref{subsec-gmc}) of a system and can find application, for instance, in the study of quantum phase transitions in critical systems
\cite{Sachdev}.

Based on the discussion about classically-correlated states and of the quantum correlation present in non-classical states, addressed in the introductory section, we propose the following definition.

\textit{Definition 1.} An $n$-partite state
$\rho_{1\cdots n}$ has genuine $n$-partite quantum correlation only
if it is not a classical state under any bipartite cut of the system,
viz., $\rho_{1\cdots n}\ne\chi_{c_{1}c_{2}}^{\rho},$ with $\chi_{c_{1}c_{2}}^{\rho}=\sum_{ic_{1},ic_{2}}p_{ic_{1},ic_{2}}|\psi_{ic_{1}}\psi_{ic_{2}}\rangle\langle\psi_{ic_{1}}\psi_{ic_{2}}|,$
where $\{p_{ic_{1},ic_{2}}=\langle\psi_{ic_{1}}\psi_{ic_{2}}|\rho_{1\cdots n}|\psi_{ic_{1}}\psi_{ic_{2}}\rangle\}$
is a probability distribution and $\{|\psi_{ic_{s}}\rangle\}$ is
an orthonormal basis for $\mathcal{H}_{c_{s}}$. 

From this definition, a quantifier of \textit{genuine $n$-partite quantum correlation} of an $n$-partite state follows as
\begin{equation}
Q_{n}(\rho_{1\cdots n}):=\min_{(c_{1},c_{2})}S(\rho_{1\cdots n}||\chi_{c_{1}c_{2}}^{\rho}).
\end{equation}
$Q_{n}(\rho_{1\cdots n})$ measures the minimum distance between $\rho_{1\cdots n}$ and the bipartite classical states $\chi_{c_{1}c_{2}}^{\rho}$. By definition, $Q_{n}(\rho_{1\cdots n})\ge0$ with equality only in cases where $\rho_{1\cdots n}=\chi_{c_{1}c_{2}}^{\rho}$. 

Now a measure for \textit{genuine $k$-partite quantum correlation} of an $n$-partite system
is defined in an analogous manner to the genuine $k$-partite total correlation introduced above, namely:
\begin{equation}
Q_{k}(\rho_{1\cdots n}):=\max_{\rho_{k}}Q_{k}(\rho_{k}).
\end{equation}
Using this quantifier, we define the \textit{degree of quantumness}, or degree of quantum correlation, of an $n$-partite system as the maximum number $k$ of its subsystems that presents genuine $k$-partite quantum correlation.

One possible definition for genuine $n$-partite classical correlation
of an $n$-partite state is given as follows. 

\textit{Definition 2.} An $n$-partite state
$\rho_{1\cdots n}$ possesses genuine $n$-partite classical correlation
only if its closest $n$-partite classical state $\chi_{1\cdots n}^{\rho}$
has genuine $n$-partite total correlation, namely, if the classical
probability distribution $p_{i1,\cdots,in}$ does not factorizes under
any bipartite cut of $\chi_{1\cdots n}^{\rho}$, i.e., $p_{i1,\cdots,in}\ne p_{c_{1}}p_{c_{2}}.$

Based on this definition, we propose the following quantifiers for
\textit{genuine $n$-partite classical correlation}:
\begin{equation}
C_{n}(\rho_{1\cdots n}):=I_{n}(\chi_{1\cdots n}^{\rho}),
\end{equation}
and for \textit{genuine $k$-partite classical correlation}:
\begin{equation}
C_{k}(\rho_{1\cdots n})=\max_{\chi_{k}^{\rho}}I_{k}(\chi_{k}^{\rho}),
\end{equation}
of an $n$-partite state. The states $\chi_{k}^{\rho}$ in the last
equation are obtained by tracing out all but the chosen $k$ subsystems
of $\chi_{1\cdots n}^{\rho}$ and the maximization is made over all
possibilities for $\chi_{k}^{\rho}$. 

Now, using $C_{k}$, we can define the \textit{degree of classical correlation} of an $n$-partite state as the maximum number $k$ of subsystems possessing genuine $k$-partite classical correlation.

In the next section we will
apply some of the quantifiers of genuine $n$-partite correlations
introduced here to study the generation of genuine multipartite system-environment correlations
in some decoherent dynamics.

\section{System-Environment Correlations in Decoherent Dynamics}

\label{sec:dynamics}

Let us consider a two-qubit system initially prepared in a Werner's state
\begin{equation}
\rho_{ab}^{w}=(1-c)\mathbb{I}_{ab}/4+c|\psi_{ab}^{-}\rangle\langle\psi_{ab}^{-}|,
\end{equation}
where $|\psi_{ab}^{-}\rangle=(|0_{a}1_{b}\rangle-|1_{a}0_{b}\rangle)/\sqrt{2}$ and $0\le c\le1$. This state has a rich structure with respect to correlations. It violates the CHSH inequality \cite{CHSH} for $c\ge 1/2$, it violates the Peres' criterion for separability \cite{PeresC} when $c> 1/3$, and it has nonzero quantum discord \cite{Ollivier-QD} for all $c\ne 0$. So, any possible qualitative difference in the multipartite system-environment correlations due to system's initial correlations (i.e., if the system state is nonlocal, nonseparable, discordant or classical) would be observed using the Werner's state as the system's initial state. 

Now these two sub-systems are let to interact locally with two independent
environments in the vacuum state $|0_{E_{s}}\rangle$, where $s=a,b$.
So, the initial state of the whole system is
\begin{equation}
\rho_{abE_{a}E_{b}}=\rho_{ab}^{w}\otimes|0_{E_{a}}\rangle\langle0_{E_{a}}|\otimes|0_{E_{b}}\rangle\langle0_{E_{b}}|.
\end{equation}
For this system, it was shown in Ref. \cite{Maziero-PRA} that, in
contrast to dissipative interactions, the decoherent dynamics of the
two qubits under phase-damping or Pauli channels does not generate
entanglement between the systems and its respective environments or
between the two environments. The bipartite quantum discord created
in such a dynamics was then indicated as the mechanism for the leakage
of quantum information out of the systems. Moreover, the initial quantum
correlation between the two qubits was shown to be not transfered
to the environments, seeming to evaporate. 

Here we extend these results
by studying the generation of genuine multipartite system-environment
correlations. We show that the quantum correlations initially shared
between the qubits do not disappear, but are transformed into genuine
multipartite correlations between systems and environments. Furthermore,
these correlations present an interesting dynamics with sudden changes
in behavior. For more results related to the sudden-change phenomenon
see Refs. \cite{Maziero-SC,Auccaise-SC,Xu-SC-Nature,Mazzola-SC,Cornelio-SC}.
Further works considering the dynamics of system-environment correlations
can be found in Refs. \cite{Lopez-ESB,Zhang-SE,Pernice-SE,Luo-Dec-Cap,Man-SE,Fanchini-SE}.

\subsection{Amplitude-damping channels}

We begin by studying the situation in which the two qubits, $a$ and
$b$, evolve under the influence of local-independent amplitude-damping
channels. The Kraus' operators for a dissipative reservoir at zero
temperature are $K_{0}=|0_{s}\rangle\langle0_{s}|+\sqrt{1-p_{s}}|1_{s}\rangle\langle1_{s}|$
and $K_{1}=\sqrt{p_{s}}|0_{s}\rangle\langle1_{s}|$, where $p_{s}$
is a parametrization of time for the subsystem $s$, with $p=0$ corresponding
to $t=0$ and $p=1$ being equivalent to $t\rightarrow\infty$ \cite{Salles-PRA}.
Throughout this article we consider identical environments and consequently
$p_{a}=p_{b}:=p$. Thus, by using the Kraus operators shown above,
we obtain the dynamical map for the system-environment evolution:
\begin{eqnarray}
U_{sE_{s}}|0_{s}0_{E_{s}}\rangle & = & |0_{s}0_{E_{s}}\rangle,\\
U_{sE_{s}}|1_{s}0_{E_{s}}\rangle & = & \sqrt{1-p_{s}}|1_{s}0_{E_{s}}\rangle+\sqrt{p_{s}}|0_{s}1_{E_{s}}\rangle.\nonumber 
\end{eqnarray}
Utilizing this map, we find the evolved state for the whole system:
\begin{equation}
\rho_{abE_{a}E_{b}}^{ad}(p)=(1-c)\iota_{ad}(p)+c|\Upsilon_{ad}(p)\rangle\langle\Upsilon_{ad}(p)|,\label{eq:Global-Evol-state}
\end{equation}
with $\iota_{ad}(p)$ and $|\Upsilon_{ad}(p)\rangle$ given in Appendix
\ref{appendix-A}. 

From $\rho_{abE_{a}E_{b}}^{ad}(p)$ we can access
the state of any partition of the system and calculate numerically
its entropy and correlations. 
\begin{figure}
\includegraphics[scale=0.37]{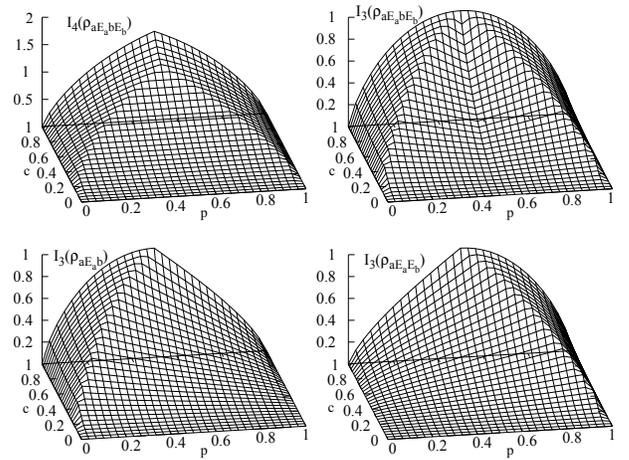}

\caption{Genuine $4$- and $3$-partite total correlation of the whole system
$aE_{a}bE_{b}$ and genuine $3$-partite correlation of $abE_{a}$
and $aE_{a}E_{b}$ for local-dissipative environments. We see that,
in general, the multiparticle correlations increase with $p$ up to
a certain instant of time from which they start to diminish going
to zero in the asymptotic limit $p=1$.}

\label{gtc-ad}
\end{figure}
\begin{figure}
\includegraphics[scale=0.52]{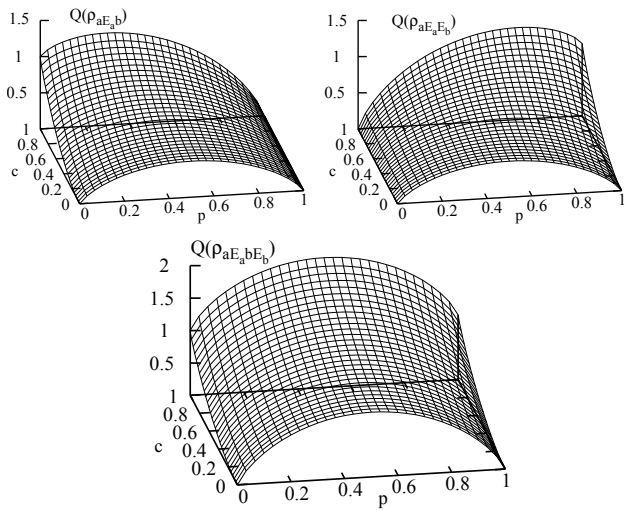}

\caption{Multipartite system-environment quantum correlations for local amplitude-damping
channels. For $p=1$ the systems state is $|0_{s}\rangle$ and its
initial correlations were altogether transfered to the environments.}

\label{qc-ad}
\end{figure}
\begin{figure}
\includegraphics[scale=0.39]{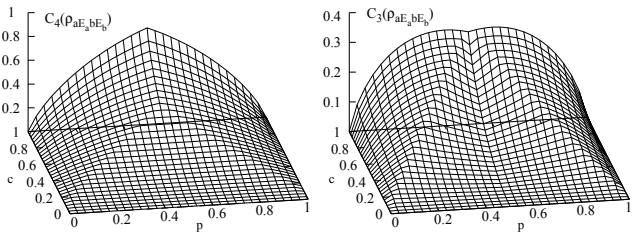}

\caption{Genuine multipartite system-environment classical correlation for
local amplitude-damping channels. The generic behavior of these correlations
is similar to that of genuine multipartite total correlation.}

\label{gcc-ad}
\end{figure}
 Here we will concentrate on multipartite correlations. A detailed
analysis of the bipartite correlations can be found in Ref. \cite{Maziero-PRA}.

We observe that, due to the assumed symmetries of the systems and
environments, the following partitions are equivalent in what refer
to correlations: $aE_{a}\equiv bE_{b}$, $aE_{b}\equiv bE_{a}$, $abE_{a}\equiv abE_{b}$,
and $aE_{a}E_{b}\equiv bE_{a}E_{b}$. Considering these symmetries we have,
from the definition of genuine $n$-partite total correlation introduced
in Section II, that
\begin{equation}
I_{4}(\rho_{aE_{a}bE_{b}})=\min_{(c_{1},c_{2})}(S(\rho_{c_{1}})+S(\rho_{c_{2}}))-S(\rho_{aE_{a}bE_{b}}^{ad}),
\end{equation}
with the following possible bi-partitions of the system: $(c_{1},c_{2})=(a,E_{a}bE_{b})$,$(E_{a},abE_{b})$,$(ab,E_{a}E_{b})$,$(aE_{a},bE_{b})$,
$(aE_{b},bE_{a})$. In the last equation $S(\rho)=-\mathrm{tr}(\rho\log_{2}\rho)$ is the von Neumann entropy. 

The $3$-partite genuine total correlation of
the whole system is given by
\begin{equation}
I_{3}(\rho_{aE_{a}bE_{b}}^{ad})=\max(I_{3}(\rho_{aE_{a}b}^{ad}),I_{3}(\rho_{aE_{a}E_{b}}^{ad}))
\end{equation}
with analogous bi-partitions used to compute the genuine $3$-partite
correlation of the $3$-partite states. All genuine total correlations
discussed above are shown in Fig. \ref{gtc-ad}. We see that, in general,
multipartite correlations are generated during the dissipative interaction
between the qubits and its respective reservoirs up to a certain instant
of time from which such correlations \textit{suddenly} begin to decrease going
to zero in the asymptotic time $p=1$. 

We also calculated the $3$-
and $4$-partite quantum correlations and genuine classical correlations
of the system, which are presented in Figs. \ref{qc-ad} and \ref{gcc-ad},
respectively. As expected, the asymptotic behavior of the genuine
3- and 4-partite classical correlations is similar to that of genuine
multipartite total correlation. We observe that the quantum correlation
remaining at $p=1$ is due solely to the environments, once the systems
state in this limit is $|0_{s}\rangle$. 

It is worthwhile mentioning that,
although explicit parametrizations for states and unitary operators
of systems with dimension greater than two are possible in principle
\cite{Bruning-Par}, all the important aspects we want to emphasize
here can be addressed without computing the genuine multipartite quantum
correlations. We leave related issues for future investigations.

\subsection{Phase-damping channels}

\begin{figure}
\includegraphics[scale=0.37]{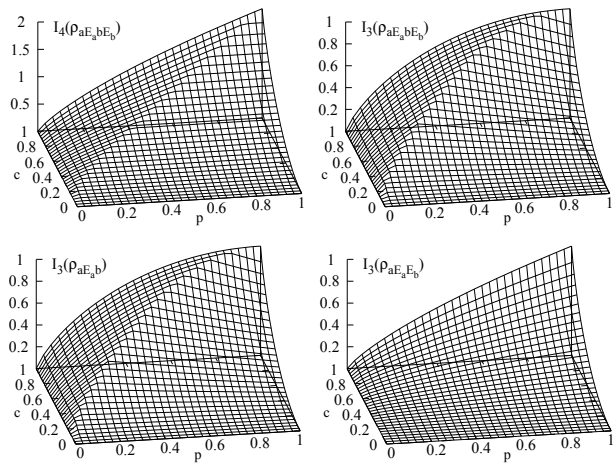}

\caption{Genuine $4$- and $3$-partite total correlation of the whole system
$aE_{a}bE_{b}$ and genuine $3$-partite correlation of $abE_{a}$
and $aE_{a}E_{b}$ for local phase-damping channels. In this case,
the genuine multipartite correlations generally increase with time
and, in contrast to amplitude-damping channels, there exists a considerable
amount of genuine $n$-partite total correlation left over in the
asymptotic limit $p=1$, depending on the purity of the initial two-qubit
state.}

\label{gtc-pd}
\end{figure}

Let us consider the dynamics of two qubits under local phase-damping
channels. This kind of noise environment causes loss of phase relations
in the system without exchange of energy. The Kraus' operators for
these channels are given by: $K_{0}=|0_{s}\rangle\langle0_{s}|+\sqrt{1-p}|1_{s}\rangle\langle1_{s}|$
and $K_{1}=\sqrt{p}|1_{s}\rangle\langle1_{s}|$. Thus, the following
map for the system-environment evolution is obtained
\begin{eqnarray}
U_{sE}|0_{s}0_{E_{s}}\rangle & = & |0_{s}0_{E_{s}}\rangle,\\
U_{sE}|1_{s}0_{E_{s}}\rangle & = & |1_{s}\rangle\otimes\left(\sqrt{1-p}|0_{E_{s}}\rangle+\sqrt{p}|1_{E_{s}}\rangle\right).\nonumber 
\end{eqnarray}
In a correspondent manner as we did for amplitude-damping channels,
we use the map shown in the last equation to compute the global evolved
state and then calculate its correlations. In this case, $\rho_{aE_{a}bE_{b}}^{pd}(p)$
is given as in Eq. (\ref{eq:Global-Evol-state}) but with $\iota_{pd}(p)$
and $|\Upsilon_{pd}(p)\rangle$ presented in Appendix \ref{Appendix-B}.
The genuine total correlations generated in the evolution under local
phase environments are shown in Fig. \ref{gtc-pd}. These correlations also exhibit the sudden-change phenomenon.

\begin{figure}
\includegraphics[scale=0.52]{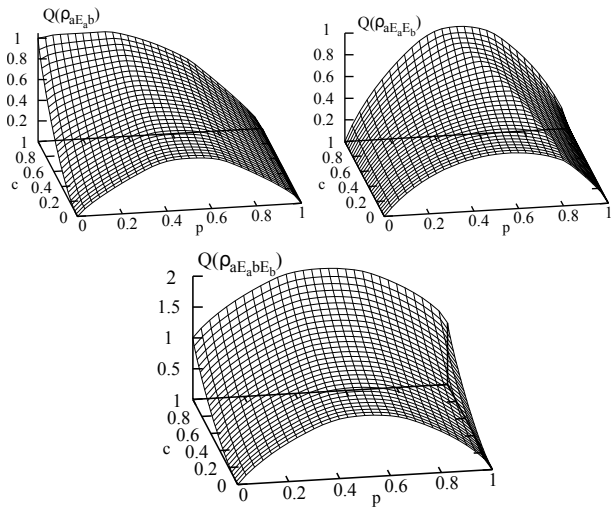}

\caption{Dynamics of $3$- and $4$-partite system-environment quantum correlations
for local phase-damping channels. Tripartite quantum correlations
are generated during the system's evolution but decay to zero at $p=1$
while there still exists multipartite quantum correlation involving
the whole system in this asymptotic limit. }

\label{qc-pd}
\end{figure}
\begin{figure}
\includegraphics[scale=0.39]{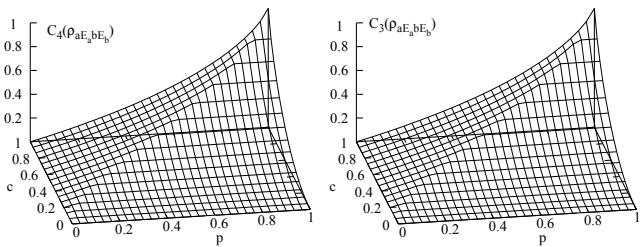}

\caption{Genuine $3$- and $4$-partite system-environment classical correlation
for local phase-damping channels. By comparing these results with
the genuine total correlations in Fig. \ref{gtc-pd}, we see that
the genuine $3$-partite classical and total correlations have the
same range and behavior.  Therefore there is no or feeble quantumness
in the genuine tripartite correlations of the system. However, as
can be observed in Fig. \ref{qc-pd}, there exists $4$-partite quantum
correlation that was created during the system-environment interaction.}

\label{gcc-pd}
\end{figure}
 In sharp contrast to the case of amplitude-damping channels, we see
that the dynamics induced by phase environments does generate genuine
multipartite correlations in the asymptotic limit $p=1$. In this
limit, the whole system will be correlated in a degree that is proportional
to the purity of the two-qubit initial state. 

We also calculated the
multipartite quantum correlations and the genuine $3$- and $4$-partite
classical correlations, which are presented in Figs. \ref{qc-pd}
and \ref{gcc-pd}, respectively. It is interesting that the genuine
$3$-partite classical and total correlations range from zero to one
for this kind of environment. This fact indicates that the system presents
no or small genuine $3$-partite quantum correlation. As can be seen
in Figs. \ref{gtc-pd}-\ref{gcc-pd}, this situation changes in the
case of $4$-partite correlations. In fact, we observe in Fig. \ref{qc-pd}
that tripartite quantum correlations are created during the system's
evolution but disappear in the asymptotic limit while there are $4$-partite
quantum correlation remnant at $p=1$.

\section{Concluding Remarks}

Summing up, we introduced operational definitions and quantifiers for genuine $n$- and $k$-partite total, quantum, and classical correlations of an $n$-partite state. We also used our correlation quantifiers to define the degree of correlation, the degree of quantumness, and the degree of classical correlation of a physical system.
Using these correlation measures, we showed that, in contrast to amplitude-damping
channels, for which the initial correlations between the qubits are
simply transfered to the environments, phase noise channels turn such
bipartite correlations into genuine multipartite system-environment
total, quantum, and classical correlations. 

Now, in order to get a better grasp of these results, let us look
at the kind of state generated during the evolution of the system
under different kinds of noise environment. It is straightforward
to verify that if the system initial state is $|\psi_{ab}^{-}\rangle$
(i.e., if $c=1$), the whole system state for amplitude-damping channels
and $p=1/2$ is
\begin{eqnarray}
|\Upsilon_{ad}(1/2)\rangle & = & (|0001\rangle+|0010\rangle-|0100\rangle-|1000\rangle)/2\nonumber \\
 & := & |\psi_{W}\rangle,
\end{eqnarray}
which is equivalent, modulo local rotations, to the well known four-qubit
$W$ state \cite{W-state}. For phase noise environments and $p=1$,
it follows that
\begin{eqnarray}
|\Upsilon_{pd}(1)\rangle & = & (|0011\rangle-|1100\rangle)/\sqrt{2}\nonumber \\
 & := & |\psi_{GHZ}\rangle,
\end{eqnarray}
which is, also modulo local unitaries, equivalent to a $GHZ$ state
(see Section I). 
\begin{figure}
\includegraphics[scale=0.42]{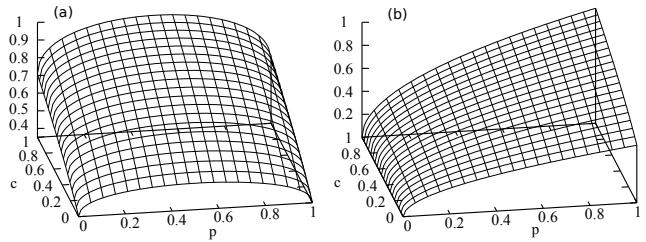}

\caption{Dependence of the fidelities (a) $F(|\psi_{W}\rangle,\rho_{aE_{a}bE_{b}}^{ad}(p))$
and (b) $F(|\psi_{GHZ}\rangle,\rho_{aE_{a}bE_{b}}^{pd}(p))$ with
$p$ and $c$. See the text for details.}

\label{fidelities}
\end{figure}
 
 For any $c$ and $p$, we obtain the following expressions for the fidelity,
\begin{equation}
F(|\psi\rangle,\rho):=\sqrt{\langle\psi|\rho|\psi\rangle},
\end{equation}
between these states and its corresponding evolved density operators:
\begin{equation}
F(|\psi_{W}\rangle,\rho_{aE_{a}bE_{b}}^{ad}(p))=\sqrt{\frac{(1+3c)(1+2\sqrt{p(1-p)})}{8}}
\end{equation}
and
\begin{equation}
F(|\psi_{GHZ}\rangle,\rho_{aE_{a}bE_{b}}^{pd}(p))=\sqrt{(1+3c)p}/2.
\end{equation}
These fidelities are shown in Fig. \ref{fidelities}.
We also observe that the mutual information is invariant under local unitary
transformations and, as one can easily verify, the genuine total correlations
for the $W$ and $GHZ$ states are given by: $I_{4}(|W\rangle)=I_{4}(|GHZ\rangle)=2$,
$I_{3}(|W\rangle)=0.81$, and $I_{3}(|GHZ\rangle)=1$. Moreover, if
we trace out a subsystem of a $GHZ$ state, the obtained tripartite
density operator has zero quantum correlations. The same action in
a $W$ state produces a $3$-partite system possessing quantumness
in its correlations. Thus, the values of the fidelities (shown in
Fig. \ref{fidelities}) and the structure of $W$ and $GHZ$ states
with respect to its correlations help us to partially explain the
general behavior of the correlations that we presented and discussed
in the last section. In reality, both kinds of decoherent dynamics generated
genuine multipartite correlations. The main difference is that for
the amplitude-damping channel these correlations are null for $p=1$
while for phase environments they generally reach its maximum value
in this limit.

Thorough investigations about the decoherence process are essential for us to obtain a better understanding of the phenomenon per se and also for the development of methods to circumvent it in the path for large scale implementations of protocols in quantum information science. From the fundamental point of view, before 2010 one believed that the flow of coherent information from the system to the environment was caused by the creation of entanglement between the two. Nevertheless, considering bipartite correlations, one of us and colleagues showed, in Ref. [57], that this is not the case in general. For some composite systems interacting with local-independent phase-damping channels (or Pauli channels), it was shown that the systems lose its initial coherent phase relations but only bipartite non-classical correlation of separable states are generated during such decoherent dynamics. Another interesting finding reported in this article was the fact that the initial bipartite quantum correlation between the 
systems (non-local, non-separable, and discordant) evaporated, i.e., only bipartite classical correlations was present at the asymptotic time of evolution. In the present manuscript, besides proposing definitions and introducing quantifiers for genuine multipartite correlations, we studied the many-body correlations generated for some important decoherent processes. Our investigation extended the previous ones in several directions, helping us to understand the global structure of the states generated during these evolutions and also explaining why the bipartite quantum correlations disappeared in the asymptotic time, by showing that they are transformed into genuine multipartite correlations. Moreover, we showed that the genuine multipartite total correlation may also exhibits a sudden change in its evolution rate.

As continuation of the present work, besides the actual calculation of the genuine multipartite quantum correlations, other interesting topic for future research is considering the dynamics of such correlations for the composition of both phase and (generalized) amplitude channels, which is a common situation in nature \cite{Auccaise-SC,Diogo-Dec-Control}.
In this case, the global state generated during the evolution under local environments will be a mixture involving W and GHZ components. The extension of these results for global and correlated environments  \cite{Diogo-quad} and to continuous variables systems \cite{Eisert-Plenio} is also worth pursuing. 

It is also important to mention that the first experiment with complete
tomography of the environment's state has been successfully performed
recently using an optical system \cite{Tom-Env}. Therefore, all the
theoretical results presented here can be verified experimentally
with current technology.

\begin{acknowledgments}
The authors acknowledge financial support from the Brazilian funding agencies Coordena\c{c}\~ao de Aperfei\c{c}oamento de Pessoal de N\'ivel Superior, Conselho Nacional de Desenvolvimento Cient\'ifico e Tecnol\'ogico, and Funda\c{c}\~ao de Amparo \`a Pesquisa do Estado do Rio Grande do Sul. J.M. thanks Marcelo S. Sarandy and Gabriel H. Aguilar for helpful discussions. We acknowledge a referee for his (her) constructive comments.
\end{acknowledgments}
\appendix

\section{Evolved State for Local Amplitude-Damping Channels}

\label{appendix-A}

For this channels, the global evolved state is given by Eq. (\ref{eq:Global-Evol-state})
with
\begin{eqnarray}
4\iota_{ad}(p) & = & |0\rangle\langle0|+(1-p)(|2\rangle\langle2|+|8\rangle\langle8|)\nonumber \\
 &  & +p^{2}|5\rangle\langle5|+(1-p)^{2}|10\rangle\langle10|\nonumber \\
 &  & +p(1-p)(|6\rangle\langle6|+|9\rangle\langle9|)+p(|4\rangle\langle4|+|1\rangle\langle1|)\nonumber \\
 &  & +\sqrt{p(1-p)}(|4\rangle\langle8|+|1\rangle\langle2|+\mathrm{h.c.})\nonumber \\
 &  & +\sqrt{p(1-p)^{3}}((|6\rangle+|9\rangle)\langle10|+\mathrm{h.c.})\nonumber \\
 &  & +p(1-p)(|5\rangle\langle10|++|6\rangle\langle9|+\mathrm{h.c.})\nonumber \\
 &  & +\sqrt{p^{3}(1-p)}(|5\rangle(\langle6|+\langle9|)+\mathrm{h.c.})
\end{eqnarray}
and
\begin{equation}
|\Upsilon_{ad}(p)\rangle=\sqrt{\frac{1-p}{2}}(|2\rangle-|8\rangle)+\sqrt{\frac{p}{2}}(|1\rangle-|4\rangle).
\end{equation}
Above, and in the next appendix, $\mathrm{h.c.}$ refers to the Hermitian
conjugate and we use the decimal representation for the indexes of
the computational bases states.

\section{Evolved State for Local Phase-Damping Channels}

\label{Appendix-B}

For phase environments, the global evolved state is given as in Eq.
(\ref{eq:Global-Evol-state}) but with
\begin{eqnarray}
4\iota_{pd}(p) & = & |0\rangle\langle0|+(1-p)(|2\rangle\langle2|+|8\rangle\langle8|)\nonumber \\
 &  & +p(|3\rangle\langle3|+|12\rangle\langle12|)+(1-p)^{2}|10\rangle\langle10|\nonumber \\
 &  & +p(1-p)(|11\rangle\langle11|+|14\rangle\langle14|)+p^{2}|15\rangle\langle15|\nonumber \\
 &  & +\sqrt{p(1-p)}(|2\rangle\langle3|+|8\rangle\langle12|+\mathrm{h.c.})\nonumber \\
 &  & +\sqrt{p(1-p)^{3}}((|14\rangle+|11\rangle)\langle10|+\mathrm{h.c.})\nonumber \\
 &  & +p(1-p)(|11\rangle\langle14|+|10\rangle\langle15|+\mathrm{h.c.})\nonumber \\
 &  & +\sqrt{p^{3}(1-p)}(|15\rangle(\langle11|+\langle14|)+\mathrm{h.c.})
\end{eqnarray}
and
\begin{equation}
|\Upsilon_{pd}(p)\rangle=\sqrt{\frac{1-p}{2}}(|2\rangle-|8\rangle)+\sqrt{\frac{p}{2}}(|3\rangle-|12\rangle).
\end{equation}

\end{document}